\documentclass[onecollarge
]{svjour2}       
\smartqed  
\usepackage{amsmath}
\usepackage{amssymb}

\newcommand{\dd}{\hbox{d}}
\newcommand{\br}{\mathbf{r}}
\newcommand{\bp}{\mathbf{p}}
\newcommand{\bq}{\mathbf{q}}

 \journalname{Journal of Statistical Physics}

\begin{document}

\title{Gravitation in the Microcanonical Ensemble: Appropriate Scaling Leading to Extensivity and Thermalization
}

\titlerunning{Gravitation in the Microcanonical Ensemble}        

\author{Maxime Champion \and Angel Alastuey \and Thierry Dauxois \and Stefano Ruffo}


\institute{ Angel Alastuey \and Maxime Champion \and Thierry Dauxois 
 \at  Laboratoire de Physique de l'Ecole Normale Sup\'erieure de Lyon, Universit\'e de Lyon and CNRS, 46, all\'ee d'Italie, F-69007 Lyon, France,\\
\and 
Stefano Ruffo \at Dipartimento di Energetica ``Sergio Stecco" and CSDC,
Universit\`a di Firenze, CNISM and INFN, via S. Marta 3, 50139 Firenze, Italy}

\date{Version: \today}

\maketitle

\begin{abstract}
We introduce a simple model of hard spheres 
with gravitational interactions, for which we study a suitable scaling limit.  
Usual extensive properties are maintained notwithstanding the long range 
of gravitational interaction. We show that a local
thermalization spontaneously emerges within a microcanonical description of the stationary 
state. In the considered scaling limit, the resulting density profile can be 
determined in a hydrostatic approach.   
 \end{abstract}

\section{Introduction}

It is widely accepted that, in order to even attempt a statistical mechanics treatment of a
system of $N$ masses that mutually interact via Newtonian forces in three dimensions, one has to 
confine them inside a box of volume $\Lambda$ to avoid evaporation. 
Confinement in a volume is a natural requirement
also for a gas of particles interacting via short-range forces, but looks inappropriate
for a model that should describe the Universe, a system that occupies full space. However,
this seems a necessity for any thermodynamical theory of a self-gravitating system in
the absence of cosmological expansion \cite{Joyce}. 

Moreover, a regularization of the 
interaction at short distances must be introduced to prevent the divergence 
of thermodynamic potentials \cite{Thirring,Kiessling,Padmanabhan,Chavanis,Pomeau2007}. 
In the case in which the Newtonian potential is softened at short distances and masses
have no spatial extension, the minimal energy scales like $N^2$ and H-stability is violated~\cite{Ruelle}. 
A mean-field (continuum) limit can, however, be performed by scaling the coupling constant
as $1/N$, which yields an extensive energy~\cite{Messer_Spohn}. It should be remarked
that energy remains non additive and, as a consequence, one finds that different statistical ensembles
give inequivalent predictions \cite{Lebowitz,Barre,Chavanis}, leading to interesting phenomena like negative 
specific heat in the microcanonical ensemble \cite{Lyn1968,Thirring}.
If particles have instead a finite radius and cannot therefore penetrate each other, the minimal energy scales 
like $N^{5/3}$ in three dimensions, violating extensivity.

In this paper we consider a self-gravitating system confined in a box and 
made of equal mass, equal shape and equal volume particles. Particles cannot penetrate
each other as a result of a hard core short-range repulsion among them. As a 
consequence, collisions among the particles are elastic and, therefore, no aggregation 
phenomenon takes place. Particles collide elastically also with the walls of the confining
box. Being then the dynamics energy and particle number conserving, it is appropriate
to consider the system in the microcanonical ensemble. The microcanonical distribution in 
phase space is a stationary solution of Liouville equation, so that it 
indeed describes a stationary state of the system~\footnote{Use of the microcanonical distribution 
amounts to assume, roughly speaking, the hypothesis of molecular chaos leading to some 
ergodicity property. In fact, as argued in Refs.~\cite{StefanoQSS,ChavanisQSS}, 
gravitational systems may be trapped in metastable states through their dynamical evolution, while the corresponding transition times 
grow exponentially fast with respect to the number of particles. That mechanism might 
induce a breakdown of ergodicity, as suggested by results obtained from numerical  simulations~\cite{PoschThirring,chabanol}. Therefore, as far as applications to real 
astrophysical systems are concerned, the microcanonical distribution has to be handled with some care.}.
The hard core interaction prevents
the divergence of the entropy. However, because of the long-range nature of the gravitational interaction,
energy is {\em a priori} non extensive and should again scale like $N^{5/3}$. We will
introduce a specific scaling limit in which we recover the extensivity of the energy.
In that scaling limit, it turns out that fluctuations of the potential energy
are small compared to the average when $N \to \infty$. Moreover, we show that if the energy 
per particle is large enough, local thermalization occurs due to hard sphere collisions. The one-body mass
distribution is then found to obey a Boltzmann-like formula in terms of the mean-field gravitational potential, 
where the average kinetic energy plays the role of temperature. The fact that a local equilibrium
is established, combined with a suitable separation of the length scales 
associated with hard core and gravitational interactions respectively, 
allows us to justify the introduction of a hydrostatic equation which express the
balance between hard-sphere pressure and gravitational attraction.

We stress that the present scaling provides a finite mass density, while the ratio 
of the kinetic energy to the gravitational energy can take arbitrary values. Therefore, it should be 
well suited for describing a wide class of physical situations. Notice that
other types of scaling have been considered in the literature \cite{DeVega2002,Joyce}, but they 
describe more specific situations like that of an infinitely diluted system for instance. 
Furthermore, once the scaling limit has been taken, the hard-core cutoff can be removed  
providing a meaningful limit  as it would be desirable. In fact at large enough energy 
per particle we recover the well-known model of the isothermal gravitational sphere \cite{Lyn1968,Emden,Antonov}.
The situation is quite different from that studied in Ref.~\cite{Kiessling} for self-gravitating
hard spheres in contact with a heat-bath in the canonical ensemble : when the size of the hard spheres 
is sent to zero while the other parameters are kept fixed, the equilibrium state becomes a 
Gaussian in velocity and a Dirac $\delta$-peak in position, in both two and three dimensions. 
Here, in our scaling limit, collapse is avoided because the gravitational interaction energy 
between two spheres at contact vanishes.

The paper is organized as follows. 
Section~\ref{DEF} presents the model and discuss the appropriate scaling continuous limit.
The scaling properties of the potential energy are presented in Sec.~\ref{SBPE}, followed by  
the discussion of the mass distribution and the fluctuation of the potential energy.
Section~\ref{thermalization} presents how a Boltzmann-like formula for the inhomogeneous density emerges spontaneously,
emphasizing a local thermalization in the microcanonical ensemble.
The hydrostatic approach is then discussed in Sec.~\ref{hydrostaticapproach}. 
We finally present our conclusions and draw some perspectives in Sec.~\ref{conclusion}.

\section{Definitions}
\label{DEF}

\subsection{\bf Hard spheres with gravitational interactions in the microcanonical ensemble}

We consider a classical gravitational model 
made of $N$ identical hard spheres with mass $m$ and diameter $\sigma$,
enclosed in a spherical box of volume $\Lambda=4\pi R^3/3$. 
The corresponding Hamiltonian reads
\begin{equation}
H_{N} = \sum_{i=1}^N {\bp_i^2 \over 2m}  + 
{1 \over 2}\sum_{i \neq j} v(|\br_i - \br_j|) 
\label{hamiltonian}
\end{equation}
with the two-body interaction potential
\begin{equation}
v(r)=\infty \;\;\; \text{for} \;\;\; r<\sigma \;\;, \;\;  v(r)=-Gm^2/r \;\;\; 
\text{for} \;\;\; r>\sigma \; .
\label{2bodypot}
\end{equation}

We consider that the previous system is isolated and does not exchange 
energy with some thermostat. Thus, its energy is fixed and equal to some 
value $E$. We assume that the corresponding stationary state is described within the
microcanonical ensemble. The corresponding distribution of positions and momenta 
of the $N$ particles in the canonical phase space reads
\begin{equation}
f_{\rm micro}(\br_1,...,\br_N,\bp_1,...,\bp_N)=A_N \; \delta (E-H_N) \; ,
\label{fmicro}
\end{equation}
while the total number of microstates is
\begin{equation}
\Omega(E,N,\Lambda) = A_N \int_{\Lambda^N \times R^{3N}} 
\prod_i \dd^3 \br_i \dd^3 \bp_i \delta (E-H_N) \; ,
\label{number}
\end{equation}
where $A_N$ is some normalization constant, which is not relevant for our purpose.

The microcanonical distribution $f_{\rm micro}$ remains stationary 
under time evolution generated by Hamiltonian $H_N$. We assume that there 
is no other conserved quantity than energy, like the total orbital momentum 
for instance~\cite{Gross}. We stress that $\Omega(E,N,\Lambda)$ is finite for $\sigma >0$, 
while it diverges for point particles with $\sigma=0$ for $N \geq 3$
as shown in Ref.~\cite{Pomeau2007}. 

\subsection{\bf The scaling continuous limit}

Since we are interested in the properties of a system with a
large number of particles, it is useful to consider some 
limit where $N \to \infty$. For ordinary systems with 
short-range interactions, that limit is nothing but the usual 
thermodynamical limit where both the energy per particle $E/N$
and the particle density $\rho_{\rm p}=N/\Lambda$ are kept fixed. 
In that limit, the physical parameters which describe one particle are 
also kept fixed. Because of 
both the attractive and long-range natures of gravitational interaction, 
such a limit would provide a collapsed state with non-extensive 
properties. In order to describe other physical situations of interest,
other limits have been introduced in the literature, like the one 
describing an infinitely diluted system~\cite{DeVega2002}. 

Here, we want to build a scaling limit when $N \to \infty$, which describes 
an infinite continuous fluid with the usual extensive properties. 
For that purpose, we consider that the parameters which define 
the particles vary with $N$ like power laws, while the gravitational 
constant $G$ is not rescaled contrarily to 
some mean-field scalings introduced in the literature~\cite{Messer_Spohn}. 
Therefore, we set $\sigma = d_0 N^{\alpha}$ and $m =m_0 N^{\delta}$, 
while the size $R$ of the spherical box is chosen to diverge 
as $\ell_0 N^{\gamma}$ with $\gamma > 0$ so that the system indeed becomes infinitely extended.
The particle density $\rho_{\rm p}=N/\Lambda$ behaves then as $N^{1-3\gamma}$. 
Since the particle are hard spheres, it is essential to keep the packing 
fraction $\eta=\pi \rho_{\rm p}\sigma^3/6$ bounded. By imposing that $\eta$
remains constant, we find a first constraint
\begin{equation}
\label{C1}
1-3\gamma + 3 \alpha = 0 \; .
\end{equation}
Moreover, we impose that the mass density $\rho=m\rho_{\rm p}$ remains 
also constant, so a second constraint arises, 
\begin{equation}
\label{C2}
1-3\gamma + \delta = 0 \; .
\end{equation}
Eventually, the extensivity of 
the gravitational energy $GM^2/R$ with the total mass $M=Nm$, provides 
the third constraint, 
\begin{equation}
\label{C3}
2+ 2\delta -\gamma = 1 \; .
\end{equation}
The three exponents are then readily 
determined as $\alpha=-2/15$, $\delta= -2/5$ and $\gamma=1/5$. 
Thus, the power laws defining 
the required scaling limit when $N \to \infty$, denoted ${\rm SL}$, are 
\begin{equation}
R=N^{1/5} \ell_0 \;\; , \;\;
m=N^{-2/5} m_0 \;\; , \;\;  \sigma=N^{-2/15} d_0 \;\;, \;\; 
E=N \frac{Gm_0^2}{\ell_0} \; \varepsilon  
\label{scaling} 
\end{equation}
with parameters $\ell_0$, $m_0$, $d_0$ and $\varepsilon$ fixed. 

Within the above SL, particles become infinitely small and light, 
while their inner mass density remains constant.  
Their number density $\rho_{\rm p}=(3/(4\pi \ell_0^3)) N^{2/5}$ 
diverges, while the mass density $\rho=m\rho_{\rm p}=3m_0/(4\pi \ell_0^3)$ 
is indeed kept fixed, as well as the packing fraction 
$\eta=\pi \rho_{\rm p}\sigma^3/6=d_0^3/(8 \ell_0^3)$. That limit 
clearly describes an infinite continuous medium, with an extensive total 
energy $E$. The corresponding stationary state is 
controlled by two independent 
dimensionless parameters, namely $\varepsilon$ which is 
the energy per particle in units of $Gm_0^2/\ell_0$, 
and the packing fraction $\eta=d_0^3/(8 \ell_0^3)$. 
Notice that, within the considered scaling, 
the mean free path $\ell=1/(\pi \sqrt{2} \; \sigma^2 \rho_{\rm p})$ 
remains proportional to the diameter~$\sigma$ of the hard spheres, 
$\ell=\sigma/(6 \sqrt{2} \; \eta)$. Therefore, for high dilutions such that 
$\eta \ll 1$, $\ell$ is much larger than~$\sigma$.

\section{Scaling properties}
\label{SBPE}

\subsection{\bf H-stability and extensivity of potential energy}

For any allowed configuration, the potential energy
\begin{equation}
V_{N} = -{1 \over 2}\sum_{i \neq j} {Gm^2 \over |\br_i - \br_j|} 
\end{equation}
is larger than that of the collapsed configuration where the $N$ hard spheres 
make a single cluster with size $L_{\rm coll} \sim N^{1/3} \sigma $, which is 
of order $-Gm^2N^2/L_{\rm coll}$. In the scaling limit, the collapse radius $L_{\rm coll}$
diverges as $N^{1/5} d_0$ and this provides the 
classical version of H-stability
\begin{equation}
V_{N} \geq -C_{\rm HS}{Gm_0^2 \over d_0} N  \; ,
\label{Hstability}
\end{equation}
where $C_{\rm HS}$ is a positive real number entirely determined by the 
geometrical arrangement of the hard spheres at maximum packing. 
On the another hand, the potential energy should reach its maximum when all particles are 
homogeneously distributed on the spherical surface of the box. That maximum 
behaves as the gravitational energy of an empty sphere carrying the surface mass 
density $Nm/(4\pi R^2)$, so we expect the upper bound 
\begin{equation}
V_{N} \leq - \frac{G m_0^2}{2 \ell_0} N  \; .
\label{upperbound}
\end{equation} 

According to bounds (\ref{Hstability}) and (\ref{upperbound}), 
$V_N$ remains of order $N$ for any configuration. This implies 
extensive lower and upper bounds for the average potential energy 
$\langle V_{N} \rangle$, where $\langle A \rangle$ 
denotes the microcanonical average of any observable $A$ weighted 
by the distribution (\ref{fmicro}). Therefore, the nonextensive behaviours of the 
average potential energy encountered in other limits, like the usual 
thermodynamical limit for instance, do not occur in the present scaling limit. 
Accordingly, it seems quite plausible that the extensity of $\langle V_{N} \rangle$ 
is ensured here, namely 
\begin{equation}
\lim_{\rm SL}\frac{\langle V_{N} \rangle}{N} = \frac{Gm_0^2}{\ell_0} \;  u(\varepsilon,\eta )  \; ,
\label{extensivity}
\end{equation} 
where $u(\varepsilon,\eta )$, the average potential energy per particle in units of $Gm_0^2/\ell_0$, 
is a well-behaved function of the intensive dimensionless parameters $\varepsilon$ and $\eta$.
We stress that a rigorous mathematical derivation of the extensive behaviour 
(\ref{extensivity}) is beyond our scope.

\subsection{\bf Mass distributions}

For further purposes, it is useful to introduce the 
$n$-body mass distributions. Let consider first
the one-body mass distribution which reads,
\begin{equation}
\rho(\br) = m\langle \sum_{i} \delta(\br_i - \br) \rangle \; .  
\label{1body}
\end{equation} 
According to the scaling properties of the potential energy 
described above, we can reasonably expect that $\rho(\br)$ 
takes a well-defined shape in the SL. That statement can be made more 
precise as follows. For dimensional and spherical symmetry reasons, we can recast $\rho(\br)$ as 
\begin{equation}
\label{1bodydim}
\rho(\br) =\rho(r) =  \rho g_N^{(1)}(q;\varepsilon,\eta)
\end{equation}
with $q=r/R$ and where the function 
$g_N^{(1)}(q;\varepsilon,\eta)$ is dimensionless. 
Notice that an explicit expression for $g_N^{(1)}(q;\varepsilon,\eta)$ 
can be readily obtained in terms of multiple integrals upon the 
dimensionless positions $\bq_i=\br_i/R$. Now, we make the assumption that 
$g_N^{(1)}(q;\varepsilon,\eta)$ goes to some well-defined 
function $g^{(1)}(q;\varepsilon,\eta)$ when $N \to \infty$ with
parameters $q,\varepsilon,\eta$ fixed, so 
\begin{equation}
\label{1bodySL}
\lim_{\rm SL}\rho(qR) = \rho g^{(1)}(q;\varepsilon,\eta) \; .
\end{equation}
The physical content of that scaling property, as well as its limitations, 
will be discussed further. 

The two-body mass distribution is
\begin{equation}
\rho^{(2)}(\br,\br') = m^2\langle \sum_{i \neq j} 
\delta(\br_i - \br)\delta(\br_j - \br') \rangle \; ,
\label{2body}
\end{equation}
while similar definitions hold for higher order $n$-body mass distributions 
with $n \geq 3$. The scaling property (\ref{1bodySL}) can be extended to 
such mass-distributions, namely
\begin{equation}
\label{2bodySL}
\lim_{\rm SL}\rho^{(2)}(R\bq,R\bq') = \rho^2 g^{(2)}(\bq,\bq';\varepsilon,\eta) \; ,
\end{equation} 
and so on. Of course, the limitations evocated above also apply to the 
present scaling behaviours.

Notice that the extensive behaviour 
(\ref{extensivity}) is consistent with 
the above scaling assumptions for 
the mass distributions. This is easily seen by starting from 
the integral expression of the average potential energy
\begin{equation}
\langle V_{N} \rangle = -{1 \over 2} \int_{\Lambda^2} \dd^3 \br \dd^3 \br' \rho^{(2)}(\br,\br')
{G \over |\br - \br'|} \; .
\label{avepot}
\end{equation}
After making the variable changes $\br \to R \bq$,
$\br' \to R \bq'$, and using the scaling behaviour (\ref{2bodySL}) 
for $\rho^{(2)}(R\bq,R\bq')$, we find that  $\langle V_{N} \rangle/N$ 
is indeed an intensive quantity in the SL given by formula 
(\ref{extensivity}) with 
\begin{equation}
u(\varepsilon,\eta ) = -{9 \over 32 \pi^2} \int_{q \leq 1,q' \leq 1} 
\dd^3 \bq \dd^3 \bq' g^{(2)}(\bq,\bq';\varepsilon,\eta)
{1 \over |\bq - \bq'|} \; .
\label{avepotpp}
\end{equation}

\subsection{\bf Fluctuations of potential energy}

Let us now consider  the fluctuations of the potential energy around its average. 
Similarly to formula~(\ref{avepot}), 
such fluctuations can be rewritten as 
\begin{eqnarray}
\langle V_{N}^2 \rangle - [\langle V_{N} \rangle]^2 &=& {1 \over 2} \int_{\Lambda^2} \dd^3 \br \dd^3 \br' \rho^{(2)}(\br,\br'){G^2 m^2 \over |\br - \br'|^2} 
+ 
\int_{\Lambda^3} \dd^3 \br \dd^3 \br' \dd^3 \br'' \rho^{(3)}(\br,\br',\br'')
{G^2 m \over |\br - \br'||\br - \br''|}\nonumber\\
&+&{1 \over 4}\!\! \int_{\Lambda^4} \!\!\!\!\dd^3 \br \dd^3 \br' \dd^3 \br'' \dd^3 \br'''\!\!
\left[\rho^{(4)}(\br,\br',\br'',\br''') - \rho^{(2)}(\br,\br')\rho^{(2)}(\br'',\br''')\right]
{G^2 \over |\br - \br'||\br'' - \br'''|} \; ,
\label{fluctuationspot}
\end{eqnarray}
where $\rho^{(3)}(\br,\br',\br'')$ and $\rho^{(4)}(\br,\br',\br'',\br''')$ are 
the three- and four-body mass distributions. In the scaling limit, 
the first two terms in expression
(\ref{fluctuationspot}) can be estimated by replacing 
$\rho^{(2)}(\br,\br')$ and $\rho^{(3)}(\br,\br',\br'')$ by their 
scaling behaviours in terms of $g^{(2)}(\bq,\bq';\varepsilon,\eta)$ 
and $g^{(3)}(\bq,\bq',\bq'';\varepsilon,\eta)$. The first term is then found to be of order $N^0$, 
while the second one behaves as $N$. 

If we estimate the third term in expression
(\ref{fluctuationspot}) by using only the scaling 
behaviours of the involved mass distributions, we 
obtain not surprisingly a $N^2$-behaviour.
However such an estimation should overestimate the exact behaviour, 
because the involved correlations 
$\left[\rho^{(4)}(\br,\br',\br'',\br''') - \rho^{(2)}(\br,\br')\rho^{(2)}(\br'',\br''')\right]$ 
can reasonably be expected to become rather small compared to $\rho^4$, for spatial 
configurations $(\br,\br',\br'',\br''')$ describing a large part of the integration 
domain $\Lambda^4$. A precise estimation cannot be performed at this level, 
so here we only assume
that the correlation contribution (\ref{4bodycorrpot}) should grow slower than $N^2$. 

The previous simple arguments suggest that the square of fluctuations 
(\ref{fluctuationspot}) should become small compared to $N^2$ in the SL, 
\textsl{i.e.}
\begin{equation}
\langle V_{N}^2 \rangle - [\langle V_{N} \rangle]^2 \sim o(N^2) \; .
\label{fluctuations}
\end{equation}
Notice that the corresponding fluctuations for 
an ordinary system with short-range interactions at 
thermodynamical equilibrium, do satisfy the behaviour (\ref{fluctuations}) 
since they are proportional to $N$. 

\subsection{\bf Scaling decomposition ansatz}

In the following, we will have to perform averages of quantities 
involving the potential energy $V_N$. Taking into account the extensivity of 
its average, as well as the behaviour (\ref{fluctuations}) of its 
fluctuations, we propose to decompose $V_N$ as
\begin{equation}
V_{N} \rightarrow \langle V_{N} \rangle + W_N  \; ,
\label{ansatz}
\end{equation}
with $\langle V_{N} \rangle=O(N)$ and $W_N=o(N)$ in the SL. 
We assume that such decomposition ansatz holds for most probable configurations
which mainly determine the averages of interest, so it should 
provide the exact behaviours in
the cases studied further.

Notice that both the extensivity of $\langle V_{N} \rangle$ and 
the fluctuations behaviour (\ref{fluctuations}), although quite 
plausible, are not well established at this level. In fact, we will 
show \textsl{a posteriori} that the mass distributions and correlations 
inferred from the decomposition ansatz, are such that the \textsl{a priori} 
assumptions about $\langle V_{N} \rangle$ and 
$(\langle V_{N}^2 \rangle - [\langle V_{N} \rangle]^2)$ are indeed satisfied.
This provides some kind of consistency check of the derivations, but of course 
it does not constitute a proof. Moreover, in the calculations 
of some averages, subtle correlations involving $V_N$
and other observables might occur, so the above decomposition ansatz 
would no longer work, even if $\langle V_{N} \rangle$ and 
$(\langle V_{N}^2 \rangle - [\langle V_{N} \rangle]^2)$ do 
behave as expected.

\section{Emergence of local thermalization}
\label{thermalization}

In this section, we study the inhomogeneous mass density $\rho(\br)$ 
given by average (\ref{1body}) for energies per particle $\varepsilon >-1/2$. 
First, we express $\rho(\br)$ in terms of the 
gravitational potential created by all the particles except one which is fixed at position 
$\br$. Then, in the SL, we rewrite exactly the corresponding spatial average 
by exploiting both the extensivity and $\varepsilon >-1/2$.   
Applying the decomposition ansatz~(\ref{ansatz}) to that exact expression, we obtain 
a Boltzmann-like formula for $\rho(\br)$ in the SL, with some temperature 
which emerges naturally. We conclude by a few comments, in particular about the 
corresponding formula for the one-body distribution $f^{(1)}(\br,\bp)$ in phase space. 

\subsection{\bf Introduction of the gravitational potential}

As a starting point, we compute the average (\ref{1body}) with the 
microcanonical distribution $f_{\rm micro}$ given by
expression (\ref{fmicro}). Then,  
the standard integration over the momenta of the $N$ particles leads to 
\begin{equation}
\rho(\br)=B(E,N,\Lambda) \int_{\Lambda^{N-1},|\br_i - \br_j| > \sigma} \prod_{i=2}^N \dd^3 \br_i  
[E-V_N(\br,\br_2,...,\br_N)]^{3N/2-1} 
\times \theta(E-V_N(\br,\br_2,...,\br_N)) \; . 
\label{rho}
\end{equation}
In that expression, $\theta(\xi)$ is the usual Heaviside function such that 
$\theta(\xi)=1$ for $\xi >0$ and $\theta(\xi)=0$ for $\xi <0$, 
while the normalization constant $B(E,N,\Lambda)$ ensures that 
the spatial integral of $\rho(\br)$ over the box does provide the total 
mass $Nm$ of the system. The conditions $|\br_i - \br_j| > \sigma$, arising from the hard 
sphere interaction, apply to any pair of particles, in particular those including
particle $1$ fixed at $\br_1 = \br$.  

For $\varepsilon >-1/2$, it turns out that $(E-V_N(\br,\br_2,...,\br_N))$ is 
positive for any spatial configuration thanks to the upper bound 
(\ref{upperbound}) for the potential energy. Then, the Heaviside 
function can be replaced by $1$ in the expression (\ref{rho}). That simplification is 
crucial for further transformations. In particular, let us
introduce the gravitational potential $\Phi(\br|\br_2,...,\br_N)$ 
at $\br$ created by the $(N-1)$ particles located at $\br_2,...,\br_N$.  
According to the decomposition 
\begin{equation}
V_{N}(\br,\br_2,...,\br_N) = V_{N-1}(\br_2,...,\br_N)  + m \Phi(\br|\br_2,...,\br_N) \; ,
\label{decomposition}
\end{equation}
we can exactly rewrite formula (\ref{rho}) as
\begin{equation}
\rho(\br)= B(E,N,\Lambda) \int_{\Lambda^{N-1}} \dd \mu_{N-1}
\prod_{i=2}^N \theta(|\br_i - \br|/\sigma -1)
\times \left[E-V_{N-1}\right]^{3/2}\left[1-{m\Phi \over E-V_{N-1} }\right]^{3N/2-1} \; ,
\label{rhobis}
\end{equation}
where $\dd \mu_{N-1}$ denotes the unnormalized microcanonical 
measure for the spatial configurations $(\br_2,...,\br_N)$ of a system made of 
$(N-1)$ particles, enclosed in the same spherical box with volume $\Lambda$ 
and the same energy $E$ as the genuine system with $N$ particles, 
\begin{equation}
\dd \mu_{N-1} = \prod_{i=2}^N \dd^3 \br_i   \prod_{i<j} \theta(|\br_i - \br_j|/\sigma -1)
\left[E-V_{N-1}(\br_2,...,\br_N)\right]^{3(N-1)/2-1} \; . 
\label{measure}
\end{equation}
Notice that, if $\varepsilon < -1/2$, then $V_{N-1}$ and $\Phi $ are coupled through the constraint 
$E-V_{N-1}-m\Phi \geq 0$, so the formula (\ref{rhobis}) \textsl{a priori} 
no longer holds. 

\subsection{\bf Exploiting the extensivity of potential energy}

Let us consider identity
\begin{equation}
\left[1-{m\Phi \over E-V_{N-1} }\right]^{3N/2-1} =
\exp \left\{ (3N/2-1) \ln\left[1-{m\Phi \over E-V_{N-1} }\right] \right\} \; .
\label{exponential}
\end{equation}
Since $m\Phi=O(1)$ and $E-V_{N-1}=O(N)$, we can expand the logarithm in powers of
$m\Phi/(E-V_{N-1})$. After multiplication by factor $(3N/2-1)$, we see that the linear 
term provides a contribution of order $O(1)$, while higher powers provide vanishing 
contributions when $N \to \infty$. Accordingly, we obtain for any spatial configuration  
\begin{equation}
\left[1-{m\Phi \over E-V_{N-1} }\right]^{3N/2-1} \sim 
\exp \left\{ -{3Nm\Phi \over 2(E-V_{N-1}) } \right\} 
\label{exponentialbis}
\end{equation}
in the SL. Inserting that asymptotic behaviour inside   
the r.h.s. of formula (\ref{rhobis}), we infer in the SL
\begin{equation}
\rho(\br) \sim B(E,N,\Lambda) \int_{\Lambda^{N-1}} \dd \mu_{N-1}
\prod_{i=2}^N \theta(|\br_i - \br|/\sigma -1) 
\times [E-V_{N-1}]^{3/2} \exp \left\{ -{3Nm\Phi \over 2(E-V_{N-1}) } \right\}  \; .
\label{rhoter}
\end{equation}

We stress that asymptotic expression (\ref{rhoter})
is exact, provided that $\varepsilon > -1/2$. The extensivity of the potential energy 
for any spatial configuration, 
which follows from the particular scaling defined here, 
plays a crucial role in the derivation of that behaviour. Within other scalings, 
which do not preserve that remarkable extensivity property, formula (\ref{rhoter}) 
would break down. 

\subsection{\bf Boltzmann-like formula}

If we introduce the kinetic energy $K_{N-1}$ of the $(N-1)$ particles for 
a given spatial configuration, the exponential factor 
on the r.h.s. of the asymptotic expression (\ref{rhoter}), 
can be seen as some kind of Boltzmann factor. However, at this level, the 
corresponding temperature fluctuates. Here, we show that such fluctuations 
can be neglected by applying the fluctuation ansatz described in Section~\ref{SBPE}.

The integral with the measure $\dd \mu_{N-1}$ in the r.h.s. of 
formula (\ref{rhoter}) is proportional to the microcanonical average 
of the quantity  
\begin{equation}
\prod_{i=2}^N \theta(|\br_i - \br|/\sigma -1) 
[E-V_{N-1}]^{3/2} \exp \left\{ -{3Nm\Phi \over 2(E-V_{N-1}) } \right\}  \; 
\label{Q}
\end{equation}
for the system with $(N-1)$ particles. If we introduce the corresponding 
microcanonical average  
\begin{equation}
\langle V_{N-1} \rangle_{N-1} = \frac{\displaystyle\int_{\Lambda^{N-1}} \dd \mu_{N-1} V_{N-1}}
{\displaystyle\int_{\Lambda^{N-1}} \dd \mu_{N-1}}
\label{avebis}
\end{equation}
of $V_{N-1}$, we can rewrite 
\begin{equation}
E-V_{N-1} = E -\langle V_{N-1} \rangle_{N-1} - W_{N-1}  \; ,
\label{ansatzbis}
\end{equation}
where $W_{N-1}$ denotes the deviation of $V_{N-1}$ with 
respect to its average for a given spatial configuration. 
Now, according to the fluctuation ansatz, we assume that for 
the most probable configurations which provide the main 
contributions to the average of quantity (\ref{Q}), $W_{N-1}$ remains 
small compared to $N$, so 
\begin{equation}
[E-V_{N-1}]^{3/2} = [E -\langle V_{N-1} \rangle_{N-1}]^{3/2}[1+o(1)]  \; ,
\label{norm}
\end{equation}
and 
\begin{equation}
\frac{3N}{2[E-V_{N-1}]} =\frac{3N}{2[E-\langle V_{N-1} \rangle_{N-1}]} + o(1)  \; .
\label{beta}
\end{equation}
Moreover, the average potential energy $\langle V_{N-1}\rangle_{N-1}$ 
for the system with $(N-1)$ particles differs by a term 
of order $O(1)$ from its counterpart $\langle V_N\rangle$ 
for the genuine system with $N$ particles. 
Thus, if we define the microcanonical temperature 
$T(E,N,\Lambda)$ through the usual relation 
\begin{equation}
E-\langle V_{N}\rangle = \langle K_{N}\rangle = \frac{3NT}{2} \; ,
\label{temperature}
\end{equation}
we eventually find in the SL
\begin{equation}
\rho(\br) \sim B(E,N,\Lambda)\left[\frac{3NT(E,N,\Lambda)}{2}\right]^{3/2} 
\int_{\Lambda^{N-1}} \dd \mu_{N-1} 
\times \prod_{i=2}^N \theta(|\br_i - \br|/\sigma -1) 
\exp \left\{ -{m\Phi \over T(E,N,\Lambda) } \right\}  \; .
\label{rhoquarte}
\end{equation}

The presence of the usual Boltzmann factor $\exp ( -m\Phi/ T ) $ in the asymptotic  
behaviour (\ref{rhoquarte}) does not mean that the whole system is in contact with 
a thermostat. We recall that the system is isolated with a well-defined 
energy $E$, as illustrated by the presence in behaviour (\ref{rhoquarte}) of the 
microcanonical measure $\dd \mu_{N-1}$ which defines the integration over 
spatial configurations. However, the emergence of temperature $T(E,N,\Lambda)$ 
can be traced back to some thermalization at a local level 
as argued and exploited in the next section. That temperature is indeed 
an intensive positive quantity in the SL, namely $T(E,N,\Lambda)$ 
reduces to $Gm_0^2/\ell_0$ times a dimensionless function 
$T^*(\varepsilon,\eta)$ when $N \to \infty$ while $\varepsilon$ and $\eta$ are the 
parameters which remain fixed. The interpretation in terms of 
a local thermodynamical equilibrium is confirmed 
by the analysis of the one-body distribution $f^{(1)}(\br,\bp)$ in phase space. 
Starting from the corresponding microcanonical definition, and    
using again both the extensivity of the potential energy and 
the fluctuation ansatz, we find that $f^{(1)}(\br,\bp)$ can be rewritten, 
in the SL, as $\rho(\br)$ times the Maxwell distribution proportional to $\exp(-\bp^2/(2mT))$.

Eventually, we point out that the asymptotic formula (\ref{rhoquarte})
relies on the fluctuation ansatz, the validity of which might break down 
for some sets of parameters $(\varepsilon,\eta)$ which determine the 
stationary state of the infinite system. 
Formula (\ref{rhoquarte}) might then fail 
for states with large collective fluctuations, as 
discussed in the next section.

\section{Hydrostatic approach}
\label{hydrostaticapproach}
\subsection{\bf Emergence of local equilibrium and separation of scales}

The emergence of a local thermodynamical equilibrium at temperature $T$, 
can be interpreted as resulting from the collisions between the hard 
spheres~\footnote{Before the SL is taken, namely for $N$ finite, the 
gravitational interactions between close particles inside a small volume surrounding $\br$ should also 
contribute to the thermalization process.}. 
The gravitational interactions between the particles 
inside a finite volume surrounding $\br$, can be safely neglected in the 
considered scaling limit since the mass of each particle then vanishes. 
In particular, notice that the gravitational interaction energy between two spheres 
at contact, $-Gm^2/\sigma$, vanishes as $N^{-2/3}$ in the SL, so it becomes 
small compared to the thermal kinetic energy $T$ which remains finite. Thus, the local 
equilibrium is entirely determined by the hard-core interactions, namely by the local 
packing fraction $\eta(\br)= \eta \rho(\br)/\rho$.

According to the above argument, at the local level around point $\br$, gravitation 
should intervene only through the gravitational potential
\begin{equation}
\phi(\br) = -\int_{\Lambda} \dd^3 \br' \rho(\br')  
{G \over |\br' - \br|} 
\label{meanpotential}
\end{equation}
created by the whole mass distribution $\rho(\br') $, which can be viewed as some external 
potential for the corresponding local system. In the expression (\ref{rhoquarte}), this 
amounts to replace the fluctuating potential $\Phi$
by $\phi(\br)$, namely to apply some kind of fluctuation ansatz to $\Phi$ 
by assuming that its fluctuations vanish in the SL.   

Within the present picture, the typical length scales become completely separated 
in the SL. On the one hand, the density 
$\rho(\br)$ is expected to vary on a length scale of order $R$, as made
explicit through the scaling behaviour (\ref{1bodySL}). On the 
other hand, the local correlation length $\lambda(\br)$ is controlled by hard sphere 
interactions, so it takes the form \cite{Hansen,Mulero}
\begin{equation}
\lambda(\br)=\sigma \xi_{\rm HS}(\eta(\br)) \; 
\label{corrlength}
\end{equation}
where $\xi_{\rm HS}(\eta)$ is some dimensionless function of $\eta$. 
In the SL with $\bq=\br/R$ fixed, since 
$\eta(R\bq)$ goes to a finite value, $\lambda(R\bq)$ behaves as $\sigma$ 
and vanishes as $N^{-2/15}$. Thus, the density indeed varies 
on a scale infinitely larger than the local correlation length. 

\subsection{\bf The coupled hydrostatic and gravitational equations}
 
The emergence of a local thermodynamical equilibrium, 
combined to the separation of length scales, 
lead us to write down the hydrostatic equilibrium equation 
\begin{equation}
\nabla P(\br) = -\rho(\br) \nabla \phi (\br)  \;,
\label{hydrostatic}
\end{equation}
which expresses the balance between pressure and gravitational forces. Furthermore, 
in that equation, the local pressure $P(\br)$ 
reduces to that of an homogeneous gas of pure
hard spheres without gravitational interactions, at temperature $T$ 
and number density $\rho(\br)/m$, \textsl{i.e.} 
\begin{equation}
P(\br) = \frac{T \rho(\br)}{m} p_{\rm HS}(\eta \rho(\br)/\rho)  \; ,
\label{PHS}
\end{equation}
\bigskip
where $p_{\rm HS}$ is the dimensionless hard-sphere pressure 
which depends only on the local packing fraction~$\eta(\br)=\eta \rho(\br)/\rho$.

The present analysis strongly suggests that
the density profile in the SL is \textsl{a priori} 
exactly determined by the coupled equations 
(\ref{meanpotential}) and (\ref{hydrostatic}), with pressure $P$ replaced by its 
hard-sphere expression~(\ref{PHS}), together with the total mass constraint  
\begin{equation}
\int_{\Lambda} \dd^3 \br \rho(\br)  = N m \; ,
\label{totalmass}
\end{equation}
and the total energy constraint
\begin{equation}
\frac{3NT}{2} + \frac{1}{2}\int_{\Lambda} \dd^3 \br \rho(\br)\phi(\br)  = E \; .
\label{totalenergy}
\end{equation}
Notice that all those equations and constraints can be recast in terms of the 
dimensionless variable $\bq=\br/R$ and of the dimensionless density profile
$g^{(1)}(q;\varepsilon,\eta)=\lim_{\rm SL} \rho(R\bq)/\rho$. We recall that temperature~$T$ 
is not a given parameter, so the corresponding dimensionless 
temperature $T^*(\varepsilon,\eta)=T/(Gm_0^2/\ell_0) $ has to be determined 
through the resolution of the whole system of equations where 
$\varepsilon$ and $\eta$ are the fixed control parameters. 

If the gravitational long-ranged interactions can be treated as a mean-field level 
through the introduction of the average potential $\phi(\br)$, the 
hard-core short-ranged interactions play a crucial role in the determination 
of the density profile. Thus, the celebrated mean-field theory for point particles~\cite{Emden,Antonov,Lyn1968} , does not provide the 
exact density profile in the present SL. The corresponding equations 
can be retrieved from the present analysis by replacing in the hydrostatic 
equation (\ref{hydrostatic}) the hard sphere 
pressure by its ideal counterpart. After an obvious integration, this provides 
the Boltzmann expression 
\begin{equation}
\rho(\br) = C \rho  \exp \left\{ -{m \phi(\br) \over T} \right\} \; ,
\label{Boltzmann}
\end{equation}
where $C$ is some dimensionless normalization constant which remains to be determined.
The resulting mean-field coupled equations for $\rho(\br)$ and 
$\phi(\br)$ can be also recast in terms of the 
dimensionless variable $\bq=\br/R$ and of the dimensionless density profile
$\rho(R\bq)/\rho$. Now $\varepsilon$ is the sole control parameter, and it 
can be rewritten as $E/(GM^2/R)$ where the 
total mass is $M=Nm$. 
The properties inferred from that so-called model of the 
isothermal sphere have been widely studied in the literature~\cite{Chav2003}. 
We stress that the presence of the hard-core interactions should 
invalidate, even at a qualitative level, some mean-field predictions. Indeed, since the local 
packing fraction $\eta(\br)=\eta \rho(\br)/\rho$ cannot 
exceed~\cite{Hale2005,Hale2010} the maximal packing fraction $\eta_{\rm max}\simeq 0.7405...$, 
the exact density $\rho(\br)$ is bounded everywhere by $\rho \eta_{\rm max}/\eta$, while 
some mean-field features are directly related to a possible unbounded increase of the 
core density~$\rho(0)$.

\subsection{\bf Consistency checks }

The derivations which ultimately provide the coupled equations 
(\ref{meanpotential}) and (\ref{hydrostatic}), involve various
\textsl{a priori} assumptions. For consistency purposes, it is 
of course necessary to check such assumptions within the global picture which 
sustains the hydrostatic approach. 

A first assumption concerns the extensivity property (\ref{extensivity}) of $\langle V_{N} \rangle$, namely the existence of a well-defined potential energy 
per particle $u(\varepsilon,\eta )$ in the SL. Within the hydrostatic approach, 
$\langle V_{N} \rangle$ reduces to the self-gravitational energy of a sphere with 
mass density $\rho(\br)$,   
\begin{equation}
V_{\rm self} = \frac{1}{2}\int_{\Lambda} \dd^3 \br \rho(\br)\phi(\br)=
-{1 \over 2} \int_{\Lambda^2} \dd^3 \br \dd^3 \br' \rho(\br)\rho(\br')
{G \over |\br - \br'|} \; . 
\label{selfpot}
\end{equation}
Since $\rho(\br)$ does satisfy the scaling property (\ref{1bodySL}), $V_{\rm self}$ 
is indeed extensive in the SL. However, notice that the exact expression (\ref{avepot}) of 
$\langle V_{N} \rangle$ can be recast as
\begin{equation}
\langle V_{N} \rangle = V_{\rm self}+ V_{\rm corr} \; ,
\label{averagepot}
\end{equation}
\bigskip
with 
\begin{equation}
V_{\rm corr} = -{1 \over 2} \int_{\Lambda^2} \dd^3 \br \dd^3 \br' \rho^{(2,{\rm T})}(\br,\br')
{G \over |\br - \br'|} 
\label{corrpot}
\end{equation}
and the truncated mass distribution or mass correlations, 
\begin{equation}
\rho^{(2,{\rm T})}(\br,\br') = \rho^{(2)}(\br,\br') - \rho(\br) \rho(\br')\; .
\label{correlation}
\end{equation}
If the coupled equations (\ref{meanpotential}) and (\ref{hydrostatic}) 
do not give access to the two-body mass correlations, the existence of 
a local thermodynamical equilibrium entirely determined by hard sphere 
interactions allow us to introduce a simple description of such 
correlations. Because of the hard core, $\rho^{(2,{\rm T})}(\br,\br')$ 
reduces to $ -\rho(\br) \rho(\br')$ for $|\br - \br' | < \sigma$.
Moreover, we expect a decay of $\rho^{(2,{\rm T})}(\br,\br')$ over 
the hard-sphere local correlation length $\lambda(\br)$. As argued 
above, $\lambda(R\bq)$ becomes proportional to $\sigma$ in the SL. Thus, the 
contribution of the vicinity of point $\br=R\bq$ to 
\begin{equation}
\int_{\Lambda} \dd^3 \br' \rho^{(2,{\rm T})}(\br,\br')
{G \over |\br - \br'|} 
\label{corrpotbis}
\end{equation}
is of order $G \rho^2 \sigma^2=O(N^{-4/15})$
in the SL. The remaining contribution to the 
integral (\ref{corrpotbis}) of points $\br'$ such that $|\br - \br'| \gg \sigma$ 
can be roughly estimated by replacing  $\rho^{(2,{\rm T})}(\br,\br')$ by a 
constant times  $m \rho(\br)/\Lambda$ which does not depend on 
$\br'$. That spread homogeneous approximation is inspired by 
the sum rule
\begin{equation}
\int_{\Lambda} \dd^3 \br'\rho^{(2,{\rm T})}(\br,\br') =-m \rho(\br)\; ,
\label{massofn}
\end{equation}
which follows from particle conservation. The corresponding 
contribution to integral (\ref{corrpotbis}) is then of order 
$G m \rho R^2/\Lambda= O(N^{-3/5})$ which becomes small compared to that
of the region $|\br - \br'| \sim \sigma$. Accordingly, we find that 
the correlation energy (\ref{corrpot}) is of order 
$G  \rho^2 R^3 \sigma^2=O(N^{1/3})$. Thus, $\langle V_{N} \rangle$ 
is indeed extensive, and the potential energy per particle is entirely 
given by the self part $V_{\rm self}$, namely
\begin{equation}
u(\varepsilon,\eta ) = -{9 \over 32 \pi^2} \int_{q \leq 1,q' \leq 1} 
\dd^3 \bq \dd^3 \bq' g^{(1)}(\bq;\varepsilon,\eta)g^{(1)}(\bq';\varepsilon,\eta)
{1 \over |\bq - \bq'|} \; .
\label{avepotppbis}
\end{equation}

Let us consider now the fluctuations 
$(\langle V_{N}^2 \rangle - [\langle V_{N} \rangle]^2)$ of the potential energy, 
which are given by formula (\ref{fluctuationspot}). We can rewrite the first 
two terms as mean-field like terms 
\begin{equation}
{1 \over 2} \int_{\Lambda^2} \dd^3 \br \dd^3 \br' \rho(\br)\rho(\br') 
{G^2 m^2 \over |\br - \br'|^2} 
\label{fluctuationspot1}
\end{equation}
and
\begin{equation} 
\int_{\Lambda^3} \dd^3 \br \dd^3 \br' \dd^3 \br'' \rho(\br)\rho(\br')\rho(\br'')
{G^2 m \over |\br - \br'||\br - \br''|} \; ,
\label{fluctuationspot2}
\end{equation}
plus the corresponding correlation terms associated with
$[\rho^{(2)}(\br,\br')-\rho(\br)\rho(\br')]$ and 
$[\rho^{(3)}(\br,\br',\br'')-\rho(\br)\rho(\br')\rho(\br'')]$. 
Since $\rho(\br)$ does satisfy the scaling property (\ref{1bodySL}), 
the mean-field contributions (\ref{fluctuationspot1}) and~(\ref{fluctuationspot2}) 
are of order $N^0$ and $N$ respectively. The corresponding 
two- and three-body correlation contributions are 
readily estimated within a simple modelization of the truncated distributions analogous 
to the one used above for analyzing contribution (\ref{corrpot}) to the 
average $\langle V_{N} \rangle$ itself. They are found to become small compared to 
their mean-field counterparts in the SL.

It remains to estimate the contribution of the third term in expression 
(\ref{fluctuationspot}) of the fluctuations 
$(\langle V_{N}^2 \rangle - [\langle V_{N} \rangle]^2)$. 
If we define
\begin{equation} 
n_{\br,\br'}^{(2)}(\br'',\br''')=\rho^{(4)}
(\br,\br',\br'',\br''')/\rho^{(2)}(\br,\br') - \rho^{(2)}(\br'',\br''') \; ,
\label{defn2}
\end{equation}
we can rewrite that four-body correlation term as
\begin{equation}
{1 \over 4} \int_{\Lambda^2} \dd^3 \br \dd^3 \br' \rho^{(2)}(\br,\br') 
{G \over |\br - \br'|}\int_{\Lambda^2} \dd^3 \br'' \dd^3 \br'''
n_{\br,\br'}^{(2)}(\br'',\br''')
{G \over |\br'' - \br'''|} \; . 
\label{4bodycorrpot}
\end{equation}
Similarly to the case of $\rho^{(2,{\rm T})}(\br,\br')$, we expect that,
for two given points $\br$ and $\br'$, $n_{\br,\br'}^{(2)}(\br'',\br''')$ 
takes non-vanishing values for spatial configurations such that 
one or more relative distances $|\br'' - \br |$, $|\br'' - \br' |$, 
$|\br''' - \br' |$, $|\br''' - \br'' |$ is of order $\sigma$. The corresponding 
contributions to the integral
\begin{equation}
\int_{\Lambda^2} \dd^3 \br'' \dd^3 \br'''
n_{\br,\br'}^{(2)}(\br'',\br''')
{G \over |\br'' - \br'''|}  
\label{integralnbis}
\end{equation}
are then readily estimated along similar lines as above when analyzing the 
two-body correlation term~(\ref{corrpotbis}). For each of the four regions 
when one of the points $\br''$ or $\br'''$ is close to either 
$\br$ or $\br'$, we find a contribution to integral (\ref{integralnbis}) 
of order $G\rho^2\sigma^3 R^2=O(1)$. After integration over $\br$ and $\br'$, 
the corresponding contribution to the term~(\ref{4bodycorrpot}) is $O(N)$. 
All the other spatial configurations $(\br'',\br''')$ for which hard-sphere 
correlations determine $n_{\br,\br'}^{(2)}(\br'',\br''')$, 
ultimately provide contributions $o(N)$ to (\ref{4bodycorrpot}). Eventually, 
it remains to determine the contributions of regions such that all 
distances $|\br'' - \br |$, $|\br'' - \br' |$, 
$|\br''' - \br' |$, $|\br''' - \br'' |$, are large compared to $\sigma$.  
As for the case of $\rho^{(2,{\rm T})}(\br,\br')$, we again use 
a spread homogeneous approximation inspired by the sum rule  
\begin{equation}
\int_{\Lambda} \dd^3 \br'' \dd^3 \br''' n_{\br,\br'}^{(2)}(\br'',\br''') =2(3-2N)m^2 \; ,
\label{massofnbis}
\end{equation}
namely we replace $n_{\br,\br'}^{(2)}(\br'',\br''')$ by a constant
proportional to $Nm^2/\Lambda^2$. This provides a contribution of order $1$ 
to integral (\ref{integralnbis}), and ultimately a contribution $O(N)$ to 
(\ref{4bodycorrpot}). Accordingly, 
the four-body correlation term (\ref{4bodycorrpot}) is
of order $N$, so fluctuations 
$(\langle V_{N}^2 \rangle - [\langle V_{N} \rangle]^2)$ 
indeed grow slower than $N^2$, in agreement with 
the assumption (\ref{fluctuations}).

\subsection{\bf Further limitations  }

We stress that there are other implicit assumptions in the hydrostatic approach, 
which might break down, at least for some sets of values for the control parameters 
$(\varepsilon,\eta)$. In particular, if the local packing fraction exceeds 
some critical value $\eta_{\rm LC}$, the system is expected to undergo a phase transition 
from a liquid to a cristalline phase~\cite{Hansen,Mulero}. Then, the density $\rho(\br)$ would oscillate
with a spatial period of order $\sigma$ in the bulk, for $\br =R\bq$ with $|\bq| <1$. Therefore, 
the hydrostatic equation which assumes local homogeneity 
on a large scale compared to $\sigma$ should be no longer valid~\footnote{Notice that close to the wall 
of the spherical box, \textsl{i.e.} for $r$ almost equal to $R$, the mass density $\rho(r)$  
varies on a scale $\sigma$. The corresponding shape does not show in the SL of 
$\rho(qR)$ for fixed $q < 1$.}. This might occur for 
sufficiently low values of $\varepsilon$ and not too small values of $\eta$. 
On the contrary, for not too low values of $\varepsilon$, and $\eta$ sufficiently small, we can expect that 
the local packing fraction $\eta(\br)$ never exceeds the critical value $\eta_{\rm LC}$, so 
the system remains in a fluid phase everywhere as implicitly assumed in the hydrostatic 
approach.

Another important assumption in the derivations relies on the existence of 
a single most probable state which provides the leading contributions to 
the averages of the quantities of interest, while the corresponding corrections 
are associated with fluctuations which remain relatively small. Within the inferred
hydrostatic approach, this means that the solution of the coupled equations 
(\ref{meanpotential}) and (\ref{hydrostatic})
has to exist and to be unique. Such existence and uniqueness are in fact not guaranteed. 
This is well illustrated by the analysis of the point-like version of those 
equations with $\eta=0$ : there are no solutions for $\varepsilon <\varepsilon_{\rm min}\simeq -0.335$ \cite{Antonov}, 
while two and more solutions can coexist in some range 
$\varepsilon_{\rm min} < \varepsilon < \varepsilon_{\rm max}$~\cite{Chav2003}. For $\eta \neq 0$, numerical 
studies~\cite{Aron1972,Stahl1995} using approximate simple forms of the hard-sphere 
equation of state $p_{\rm HS}(\eta)$ indicate that some of the previous features 
are still observed, so they are not specific to the point-like case $\eta=0$. In fact, they 
are related to the attractive nature of gravitational interactions which favors
the collapse of the system. The
absence of solutions to the coupled equations 
(\ref{meanpotential}) and (\ref{hydrostatic}) might be related to the flatness 
of the microcanonical distribution. When two or more solutions coexist, one
would expect some kind of multimodal shape for that distribution. In 
any of such cases, there appears macroscopic fluctuations which cannot be neglected 
with respect to the averages themselves. Thus, the whole scheme 
leading to the hydrostatic approach should then break down. A proper 
many-body microcanonical analysis 
of the system, which might then undergo some phase transitions~\cite{Chav2003}, 
remains a challenging and rather difficult problem. 

The tendency to collapse induces either
the violation of the condition $\eta(\br) < \eta_{\rm LC}$ near the core $r=0$, 
or the lack of the uniqueness property of the hydrostatic solution. 
That mechanism is of course related to Jeans instability~\cite{Jeans}, 
which is controlled by the Jeans length $L_{\rm J}\propto(T/(G\rho m))^{1/2}$.
Within the SL, $L_{\rm J}$ reduces to the size $R$ of the system times 
$[T^*(\varepsilon,\eta)/3]^{1/2}$. When $\varepsilon \to \infty$ at a fixed 
$\eta < \eta_{\rm LC}$, since the potential energy is always negative, the dimensionless 
temperature $T^*(\varepsilon,\eta)$ also diverges. More precisely, the lower and upper 
bounds (\ref{Hstability}) and (\ref{upperbound}) on that potential energy, 
imply the behaviour $T^*(\varepsilon,\eta) \sim 2\varepsilon/3$ 
when $\varepsilon \to \infty$ at a fixed  $\eta < \eta_{\rm LC}$. Accordingly, 
the ratio $L_{\rm J}/R$ also diverges, so previous limitations arising from 
the tendency to collapse should not occur. Thus, for $\varepsilon$ sufficiently large 
and $\eta$ sufficiently small, the hydrostatic approach can be reasonably 
expected to provide the exact density profile. In fact, the 
gravitational interactions can then be treated as a small perturbation, so the density $\rho(\br)$
should be almost uniform and close to $\rho$, in relation with the quite 
plausible limit behaviour $g^{(1)}(\bq;\varepsilon,\eta) \to 3/(4\pi)$ when $\varepsilon \to \infty$ 
with $q$ and $\eta < \eta_{\rm LC}$ fixed. When $\varepsilon$ is decreased, 
temperature $T^*(\varepsilon,\eta)$ should decrease so the ratio 
$L_{\rm J}/R$ should also decrease. When $L_{\rm J}$ becomes of 
the order of the size $R$, or even smaller than $R$, Jeans-like 
instabilities should play an important role and invalidate the assumptions 
underlying the hydrostatic approach. Therefore, there should exist some threshold 
value $\varepsilon_{\rm c}(\eta)$ depending on $\eta$, below which 
the hydrostatic approach fails. Notice that, if the derivation of the hydrostatic 
approach has been carried out with the condition $\varepsilon > -1/2$, this does not mean that 
$\varepsilon_{\rm c}(\eta)$ is necessarily smaller than $-1/2$, as suggested by the point-like case 
$\eta=0$ for which no mean-field solutions exist for 
$\varepsilon <\varepsilon_{\rm min}\simeq -0.335$. 

\vfill\eject
\section{Conclusion}
\label{conclusion}

We have presented in this paper a novel approach to the statistical mechanics of self-gravitating particles
in the microcanonical ensemble which emphasizes that gravitational interactions can be treated at 
the mean-field level. The main idea is that, while the gravitational constant is not rescaled,
an appropriate scaling of the parameters of the particle allows us to define a non-trivial equilibrum 
state with the usual H-stability and extensivity properties. If this scaling is introduced
through simple physical requirements, it turns out that, rather unexpectedly, it 
also ensures the spontaneous emergence of the local thermodynamical equilibrium~(\ref{Boltzmann}).

Several comments and open questions are however important to draw at this stage. 

i) We show here that, locally, the gas of self-gravitating particles is thermalized through collisions,
so the present scaling defines a temperature. Moreover this finding justifies a posteriori some previous
attempts to study self-gravitating particles within a canonical ensemble~\cite{Kiessling,Chavanis}.
However, that thermalization is expected to occur only at sufficiently large enough energy 
per particle. Furthermore, although the thermalization is recovered at the local level, there is still no 
clear definition of an external thermostat at the macroscopic level,
with which the system is exchanging heat or kinetic energy. Indeed, as it is well known, 
the energy being conserved, a gain (or loss) of kinetic energy can be obtained by a decrease 
(or increase) of potential energy. It is still debated nowadays~\cite{LesHouches2002}
how one can use a canonical framework with a clear physical meaning.

ii) The hard-core regularization is essential, on the one hand for avoiding the 
collapse of the finite system \cite{Pomeau2007}, and on the other hand for 
ensuring through collisions the emergence of 
a local equilibrium in the scaling limit. Once that limit has been 
taken, hard-core effects intervene in the local pressure which ultimately determines the 
density profile, at least for a large enough energy per particle. At this level, the 
hard core can be removed and we recover the density profile for point particles determined 
within the familiar mean-field description of the isothermal sphere.

iii) We have considered throughout the paper a large enough energy per particle $\varepsilon$, 
but of course states with negative $\varepsilon$ are of strong interest. Indeed, we have discovered that
positive values prevent the Jeans length to be 
small enough and thus hinder any equilibrium collapsed state. A non vanishing 
fraction of matter will stay on the boundary of the system,
even if the number of particles $N$ diverges to infinity. Statistically, 
a large part of these particles will have momenta pointing outwards,
leading to the partial evaporation of the system. This effect is not 
described within the present model in which all particles are maintained artificially
into the spherical domain. To avoid this problem, one might consider negative energies, which would lead to sufficiently 
small Jeans length to constrain particles to stay far from the boundaries: the fraction 
of evaporating particles might therefore be vanishingly small. Interestingly,
this effect is reminiscent of two particles in a Keplerian system,
which are restricted to confined elliptic trajectories only for negative energies. 

\medskip

Both long-range and short-range effects, as well as confinement, can therefore be traced back in the results 
derived within this new scaling approach to statistical mechanics in the microcanonical ensemble. 
They are at the origin of the difficulties, but also of the interests of this fascinating problem. 
We are however convinced that this new scaling approach is an important step forward 
for understanding the complete statistical mechanics treatment 
encompassing the possibility to tackle phase transitions and fragmentations of a system of self-gravitating particles.

\begin{acknowledgements}
This work has been supported by the contract LORIS (ANR-10-CEXC-010-01).
\end{acknowledgements}

\end{document}